\documentclass[twocolumn,10pt]{article}

\usepackage[T1]{fontenc}        
\usepackage{lmodern}
\usepackage{graphicx}
\usepackage{siunitx}
\usepackage[version=3]{mhchem}                      
\DeclareSIUnit\permille{\text{\textperthousand}}
\usepackage{tikz}
\usepackage{booktabs}
\usepackage{verbatim}
\AtBeginEnvironment{thebibliography}{}

\newcommand*{\A}{$^{(a)}$}                          
\newcommand*{\B}{$^{(b)}$}

\newcommand*\circledd[1]{\tikz[baseline=(char.base)]{       
            \node[shape=circle,draw,inner sep=1pt] (char) {#1};}}

\title{\textbf{Correlating parent-fragment relationships in cluster photoionization}}
\author
{Jong Chan Lee,$^{1,2}$ Beg\"um Rukiye \"Ozer,$^{1}$ In Heo,$^{1}$ Thomas Schultz$^{1,\ast}$\\
\small{$^{1}$\emph{Dept. of Chemistry, Ulsan National Institute of Science and Technology (UNIST), 44919 Korea}}\\
\small{$^{2}$\emph{Now at: Dept. of Chemistry and Applied Biosciences, ETH Z\"urich, Switzerland}}\\
\small{$^\ast$\emph{E-mail:  schultz@unist.ac.kr.}}
}
\date{}

\begin{document}

\baselineskip13pt       
\maketitle              
\interfootnotelinepenalty=1000              
\renewcommand\dblfloatpagefraction{.90}     
\renewcommand\dbltopfraction{.80}           
\renewcommand\floatpagefraction{.90}        
\renewcommand\topfraction{.80}              
\renewcommand\bottomfraction{.80}           
\renewcommand\textfraction{.1}              

\renewcommand{\abstractname}{\vspace{-\baselineskip}}   

\begin{abstract}
\bfseries 
Fragment signals in ordinary mass spectra carry no label to identify their parent molecule. By correlating mass signals with rotational Raman spectra, we created a method to label each ion signal with the spectroscopic fingerprint of its neutral parent molecule. In data for a carbon disulfide molecular cluster beam, we assigned 28 distinct ionization and fragmentation channels based on their mass-correlated rotational fingerprints. Unexpected observations included the formation of energetic \ce{S2} and \ce{SCCS} cationic fragments from the \ce{CS2}-dimer cluster and a significant \ce{CS3} signal, uncorrelated to the dimer. The large number of observed channels revealed a surprising complexity that could only be addressed with correlated spectroscopy and computer-aided correlation analysis.
\mdseries
\end{abstract}


\section*{Introduction}
Traditional spectroscopic methods are ill-suited for the investigation of heterogeneous samples. Spectroscopic signals carry no label to identify the signal origin and guesswork is required to assign an observed spectroscopic signals to a specific sample component. As established in the fields of NMR \cite{Ernst1992} and vibrational or electronic spectroscopy \cite{Jonas2003a,Hochstrasser2007,Cho2008,Biswas2022,Fresch2023}, multi-dimensional (correlated) spectroscopy can resolve heterogeneity in condensed phase samples. The development of comparable gas phase experiments is an active research field, but has been limited to rather small molecular systems \cite{Roeding2018,Bruder2018,Chen2023}. Our own method of mass-correlated rotational alignment spectroscopy (CRASY) correlates two very different observables, namely the mass of photoionized molecules with high-resolution rotational Raman spectra of their neutral precursors \cite{Schroter2011,Schroter2018,Schultz2023}. In this manuscript, we describe the use of CRASY to investigate a heterogeneous molecular cluster beam and demonstrate how this technique can assign parent-fragment relationships in a highly heterogeneous sample.

CRASY is a type of rotational coherence spectroscopy \cite{Felker1992,Riehn2002,Frey2011}, based on the excitation and probing of a coherent superposition of rotational states in the time domain. Carbon disulfide, \ce{CS2}, was the first molecule investigated by an RCS-type experiment \cite{Heritage1975} and remained a model system for RCS and related molecular alignment experiments \cite{Kumarappan2008,Bisgaard2009,Yang2015,Pickering2018,Schouder2020}. In CRASY, the rotational coherence is probed by photoionization and resulting data correlates ion signals with the rotational Raman spectra of their neutral precursor molecules. The rotational spectra can serve as a molecular fingerprints to identify each molecular species and can also be used to determine precise molecular geometries \cite{Ozer2020,Heo2022a,Heo2022b}. The correlated ion signals reveal thermodynamically and kinetically accessible fragmentation channels upon photoionization. The combined information allows to map out ionization and fragmentation channels for each molecule in a molecular beam.

Isolated molecular clusters represent a particularly interesting class of heterogeneous molecular samples, because they can serve as model systems for the observation of specific intermolecular interactions and reactions \cite{Castleman1994}. Without the complex environment of the condensed phase, spectra can be measured at high resolution and are readily compared to high level theory. But neutral cluster sources emit a heterogeneous range of cluster structures and sizes and it can be astonishingly difficult to separate and assign observed signals even for apparently simple cluster systems. An illustrative example can be found in the literature on phenol ammonia clusters, where decades of contradictory claims about size-specific reaction channels resulted in dedicated review articles \cite{David2002,Jouvet2021}.



Here, we investigate \ce{CS2} molecules and clusters in a cold molecular beam, created by pulsed jet expansion of \ce{CS2}. Rotational Raman transitions were excited (pumped) by a 1 ps, 796 nm laser pulse in the spectrometer region of a Wiley-McLaren time-of-flight mass spectrometer. The resulting coherent superposition of rotational states (wavepacket) was probed at different time delays by resonant two-photon ionization with a 50 fs, 200 nm laser pulse. As the wavepacket evolved, the probed rotational states went through moments of constructive and destructive quantum state interference. This modulated the detected ion signal amplitudes and we observed delay-dependent signal oscillations for each ion mass. These oscillations were mapped by performing 25'000 pump-probe experiments, scanning the pump-probe delay with 2.5 ps step size over a delay range of 50 ns. The Fourier transformation of signal oscillations revealed mass-correlated rotational spectra. Because the coherent rotation was excited and probed in neutral ground state molecules before photoionization, our data correlated each observed ion species with the rotational spectrum of its neutral parent molecule.

\begin{figure*}[htb]
\centering
\includegraphics[width=11.2cm]{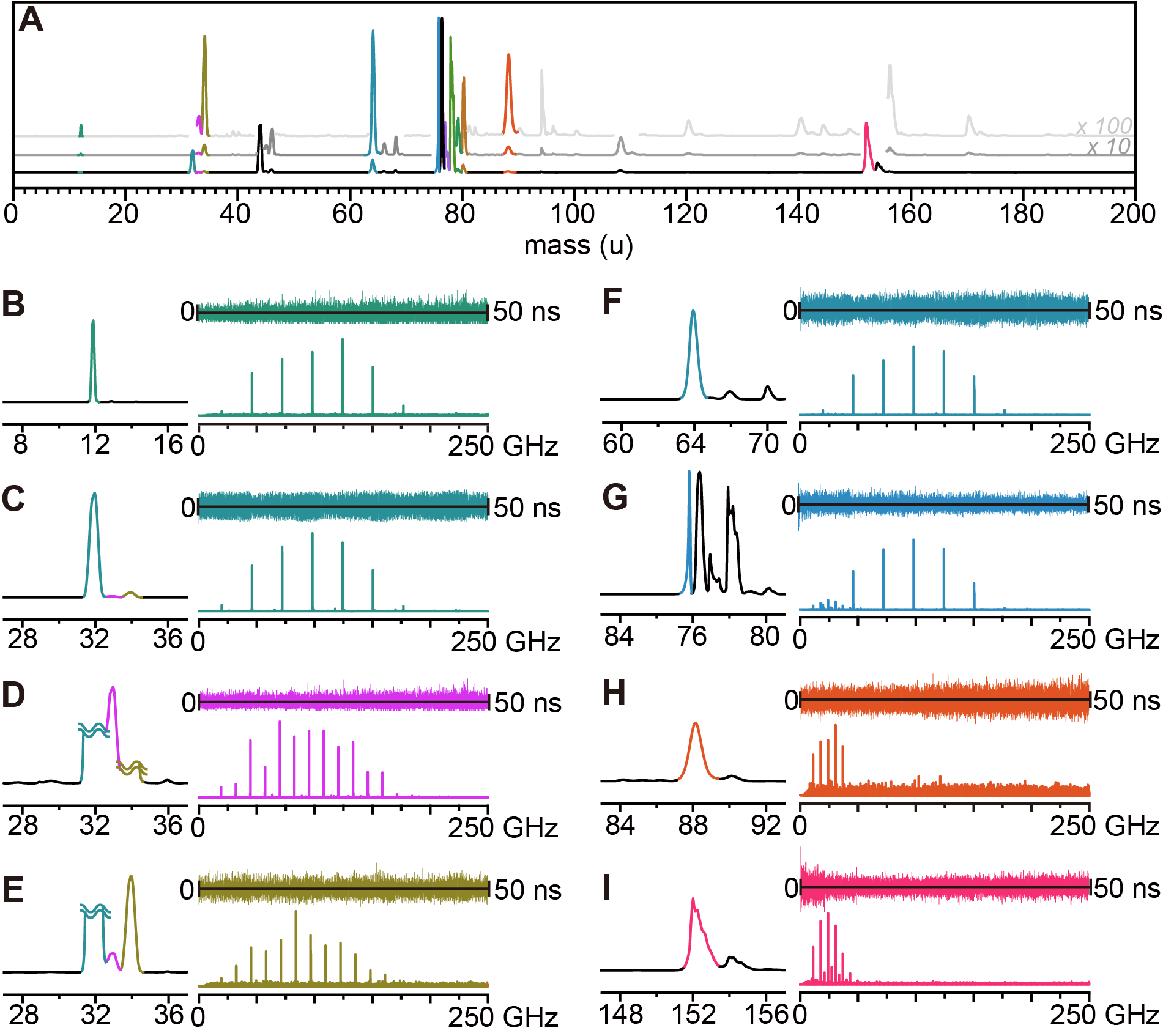}
  \caption{\small (A) Mass spectrum of \ce{CS2} isotopologues, clusters, and fragments. Grey traces with vertical offset show the same spectrum with enlarged ordinate. (B-I) Enlarged sections of the mass spectrum, delay-dependent signal modulation traces for the highlighted masses, and corresponding rotational Raman spectra. Color is used to highlight selected ion signals and is consistent across all manuscript figures and tables. }
\label{fgr:CRASY_DATA}
\end{figure*}

Cluster mass spectra are shown in Fig.\ \ref{fgr:CRASY_DATA}A and revealed 24 major ion signals with a signal amplitude exceeding \SI{1}{\permille} of the \ce{CS2} main isotopologue signal. Consistent colors are used to highlight selected mass-correlated signals in all manuscript figures and tables. Large signals were observed for abundant \ce{CS2} isotopologues (mass 76--80 u), isotopologue fragments \ce{C} (12 u), \ce{S} (32--34 u), \ce{CS} (44--46 u), \ce{S2} (64--66 u) and the \ce{CS2} dimer (152 u). The most abundant \ce{CS2} monomer isotopologue signals were saturated and showed reduced amplitudes and distorted signal shapes. Ion signals for rare isotopes were detected at their natural abundance \cite{Michael2011}, i.e., 1.1\% for \ce{^{13}C}, 0.75\% for \ce{^{33}S}, and 4.2\% for \ce{^{34}S}. Signal amplitudes for \ce{CS3} (108 u), \ce{C2S2} (88 u), and \ce{(CS2)3} (228 u) were small but detectable, as were signals for sample impurities and other \ce{C_nS_m} species.

Figs.\ \ref{fgr:CRASY_DATA} B to I show ion signals, correlated signal modulation traces, and rotational Raman spectra for a few selected mass channels. Pronounced signal oscillations were observed for all major ion signals and delay dependent signal modulation traces were calculated by summing ion signals within the colored integration boundaries. Traces were Fourier-transformed to obtain mass-correlated rotational Raman spectra. Similar spectra were observed in the mass channels for carbon (12 u, Fig.\ \ref{fgr:CRASY_DATA}B), sulfur (32 u, Fig.\ \ref{fgr:CRASY_DATA}C), \ce{CS} (44 u), \ce{S2} (64 u, Fig.\ \ref{fgr:CRASY_DATA}F), and \ce{CS2} (76 u, Fig.\ \ref{fgr:CRASY_DATA}G), indicating a common parent for these ions. Different spectra were observed for heavy isotopologues and their fragments, e.g., as plotted for the \ce{^{33}S} and \ce{^{34}S} mass channels (Figs.\ \ref{fgr:CRASY_DATA} D,E). Isotopologue rotational constants derived from the \ce{CS2} isotopologue rotational spectra agreed well with those in the scientific literature \cite{Winther1988,Cheng1996,Ahonen1997,Horneman2005,Schroter2011,Schroter2018}. 

Raman transitions for the \ce{CS2} dimer (mass 152 u, Fig.\ \ref{fgr:CRASY_DATA}I) were resolved in the frequency region $<$50 GHz. The spectrum of the dimer was previously characterized by IR spectroscopy\cite{Rezaei2011,Barclay2019} and RCS experiments\cite{Chatterley2020,Schouder2022} and observed line positions agreed with those expected from literature rotational constants. Lines from the dimer spectrum also appeared in the fragment channels of \ce{C2S2} (88 u, Fig.\ \ref{fgr:CRASY_DATA}H), \ce{CS2} (76 u, Fig.\ \ref{fgr:CRASY_DATA}G), and \ce{S2} (64 u, Fig.\ \ref{fgr:CRASY_DATA}F). 

\begin{figure*}[htb]
\centering
\includegraphics[width=13.2cm]{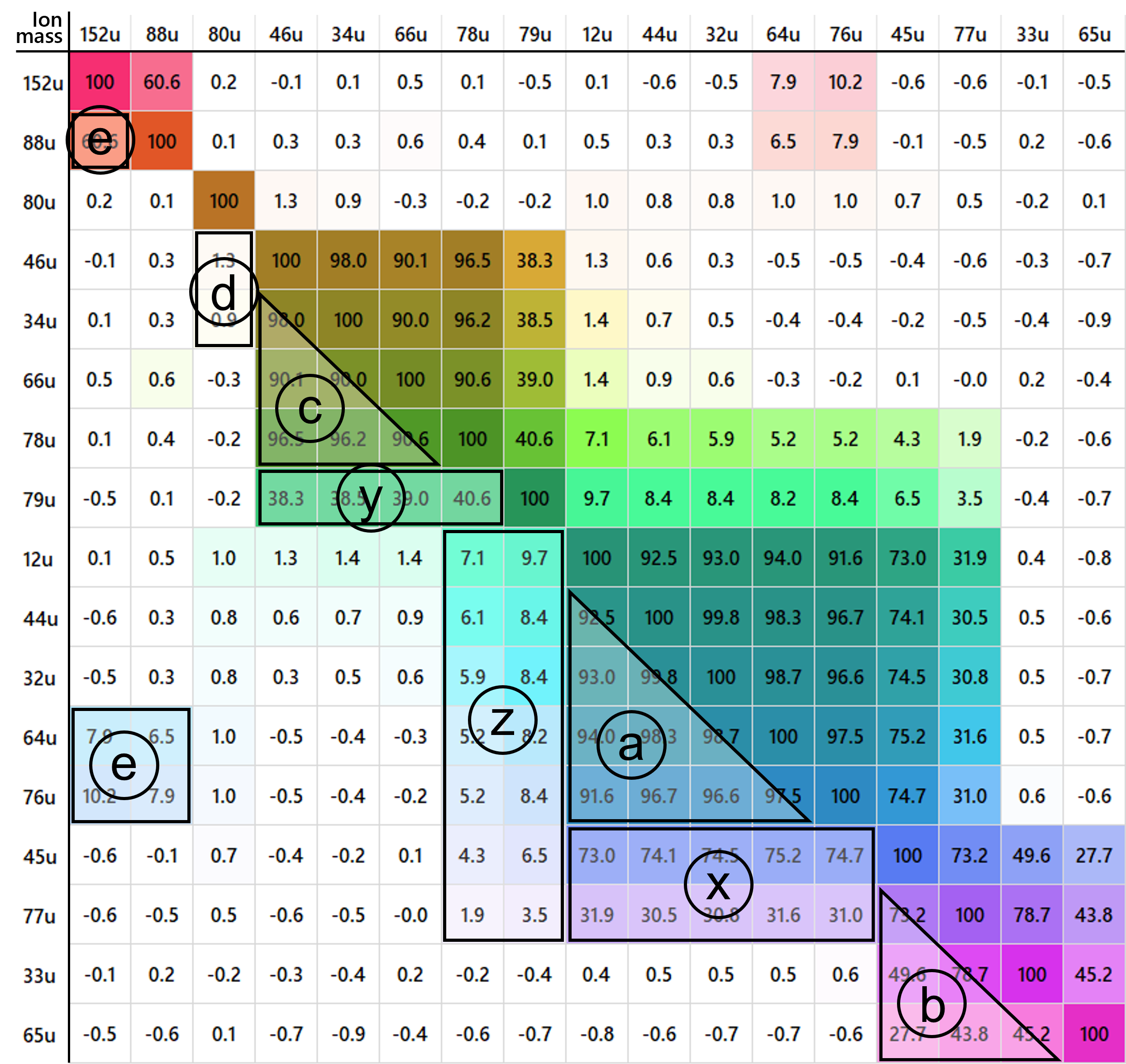}
  \caption{\small Pearson correlation coefficients for mass correlated rotational spectra, scaled to values of $\pm100$. Highlighted blocks of correlated signals are discussed in the text. Column colors correspond to those used for corresponding masses in Figs.\ 1, 3, and Table 1.}
\label{fgr:CorrelationTable}
\end{figure*}

Due to the large data quantity, a systematic analysis of parent-fragment correlations required automated analysis routines. The Pearson correlation coefficient offers a numerical measure for the similarity of spectra and Fig.\ \ref{fgr:CorrelationTable} plots correlation coefficients between major mass channels. A correlation value of 100 indicates identical spectra, 0 indicates absence of correlation, and negative values indicate anti-correlation. To keep the table at a reasonable size, signals without significant correlation to \ce{CS2} species were omitted.

For ion signals with significant correlation, we reviewed molecule-specific rotational line positions to demonstrate specific parent-fragment relationships. Fragmentation channels for confirmed relationships are summarized in Table \ref{tab:FragmentationTable_1}. The review of line positions was required to exclude spurious correlations that might occur for three reasons: \emph{(i)} Isotopologue pairs that differed only in the central \ce{^{12}C} or \ce{^{13}C} isotope showed identical or near-identical rotational transition line frequencies.\footnote{The carbon atom position coincides with the molecular center-of-mass mass and has a minute effect on the molecular inertial moment and rotational transition frequencies.} In these cases, we only assigned parent-fragment correlations for the most abundant isotopologue. \emph{(ii)} Several mass channels contained signals for multiple isotopologues. In these cases, isotope-specific lines were tracked to establish parent-fragment relationships for each isotopologue. \emph{(iii)} Due to enormous differences in isotopologue signal amplitudes and the limited dynamic range and resolution of our mass spectrometer, spectral signatures leaked from the strongest \ce{CS2} isotopologue mass channels to those of adjacent weak isotopologues. We minimize this spectral contamination by shifting signal integration boundaries towards lower mass (for 76 u) or higher mass (for 77 u, 78 u, 79 u and 80 u) and excluded lines that might stem from spectral leakage from our analysis.

\newcommand{\cb}[2]{\color{white}\colorbox{#1}{\textbf{\textsf{#2}}}\color{black}}
\definecolor{c152}{HTML}{f62f72}
\definecolor{c88}{HTML}{e15323}
\definecolor{c80}{HTML}{b97326}
\definecolor{c46}{HTML}{a27e26}
\definecolor{c34}{HTML}{8e8526}
\definecolor{c66}{HTML}{768c26}
\definecolor{c78}{HTML}{4d9426}
\definecolor{c79}{HTML}{279559}
\definecolor{c12}{HTML}{289376}
\definecolor{c44}{HTML}{299288}
\definecolor{c32}{HTML}{2a9097}
\definecolor{c64}{HTML}{2b8ea8}
\definecolor{c76}{HTML}{2e8ac3}
\definecolor{c45}{HTML}{587bf2}
\definecolor{c77}{HTML}{a461f2}
\definecolor{c33}{HTML}{d732ec}
\definecolor{c65}{HTML}{e52fc7}
\begin{table}[htb!]
\small
   \caption{Assigned photoionization and fragmentation channels.}
    \setlength\tabcolsep{4.5pt}         
    \renewcommand{\arraystretch}{0.7}   
\centering
 \begin{minipage}{7.4cm}
  \resizebox{7.4cm}{!}{
   \begin{tabular}{ll@{\hspace{1\tabcolsep}}l} \\                                          
    Neutral & Ion              &   Fragmentation channel                                \\ 
   \toprule {\ce{CS2}}
            & \cb{c76}{76 u}   &      \ce{CS2      ->[h\nu]                   CS2^{+}} \\ 
            & \cb{c64}{64 u}   &      \ce{CS2      ->[h\nu] C          +       S2^{+}} \\ 
            & \cb{c44}{44 u}   &      \ce{CS2      ->[h\nu] S           +      CS^{+}} \\ 
            & \cb{c32}{32 u}   &      \ce{CS2      ->[h\nu] CS          +       S^{+}} \\ 
            & \cb{c12}{12 u}   &      \ce{CS2      ->[h\nu] S2          +       C^{+}} \\ 
   \midrule {\ce{^{13}CS2}}
            & \cb{c77}{77 u}   & \ce{^{13}CS2      ->[h\nu]              ^{13}CS2^{+}} \\ 
            & \cb{c45}{45 u}   & \ce{^{13}CS2      ->[h\nu] S           + ^{13}CS^{+}} \\ 
   \midrule {\ce{^{33}SCS}}
            & \cb{c77}{77 u}   & \ce{^{33}SCS      ->[h\nu]              ^{33}SCS^{+}} \\ 
            & \cb{c65}{65 u}\A & \ce{^{33}SCS      ->[h\nu] C           + ^{33}SS^{+}} \\ 
            & \cb{c45}{45 u}   & \ce{^{33}SCS      ->[h\nu] S           + ^{33}SC^{+}} \\ 
            & \cb{c33}{33 u}   & \ce{^{33}SCS      ->[h\nu] CS          +  ^{33}S^{+}} \\ 
   \midrule {\ce{^{34}SCS}}
            & \cb{c78}{78 u}   & \ce{^{34}SCS      ->[h\nu]              ^{34}SCS^{+}} \\ 
            & \cb{c66}{66 u}   & \ce{^{34}SCS      ->[h\nu] C           + ^{34}SS^{+}} \\ 
            & \cb{c46}{46 u}   & \ce{^{34}SCS      ->[h\nu] S           + ^{34}SC^{+}} \\ 
            & \cb{c44}{44 u}\A & \ce{^{34}SCS      ->[h\nu] ^{34}S      +      CS^{+}} \\ 
            & \cb{c34}{34 u}   & \ce{^{34}SCS      ->[h\nu] CS          +  ^{34}S^{+}} \\ 
            & \cb{c32}{32 u}\A & \ce{^{34}SCS      ->[h\nu] C^{34}S     +       S^{+}} \\ 
            & \cb{c12}{12 u}\A & \ce{^{34}SCS      ->[h\nu] ^{34}SS     +       C^{+}} \\ 
   \midrule {\ce{^{33}S^{33}SC}}
            & \cb{c78}{78 u}   & \ce{^{33}S^{33}SC ->[h\nu]         ^{33}S^{33}SC^{+}} \\ 
   \midrule {\ce{^{34}S^{13}CS}}
            & \cb{c79}{79 u}\B & \ce{^{34}S^{13}CS ->[h\nu]         ^{34}S^{13}CS^{+}} \\ 
   \midrule {\ce{^{34}S^{33}SC}}
            & \cb{c79}{79 u}   & \ce{^{34}S^{33}SC ->[h\nu]         ^{34}S^{33}SC^{+}} \\ 
   \midrule {\ce{^{34}S^{34}SC}}
            & \cb{c80}{80 u}   & \ce{^{34}S^{34}SC ->[h\nu]         ^{34}S^{34}SC^{+}} \\ 
            & \cb{c46}{46 u}\A & \ce{^{34}S^{34}SC ->[h\nu]      ^{34}S + ^{34}SC^{+}} \\ 
            & \cb{c34}{34 u}\A & \ce{^{34}S^{34}SC ->[h\nu] C^{34}S     +  ^{34}S^{+}} \\ 
   \midrule {\ce{(CS2)_2}}
            & \cb{c152}{152 u} & \ce{(CS2)_2       ->[h\nu]               (CS2)_2^{+}} \\ 
            & \cb{c88} {88 u}  & \ce{(CS2)_2       ->[h\nu] S2          +    C2S2^{+}} \\ 
            & \cb{c76} {76 u}  & \ce{(CS2)_2       ->[h\nu] CS2         +     CS2^{+}} \\ 
            & \cb{c64} {64 u}  & \ce{(CS2)_2       ->[h\nu] C2S2        +      S2^{+}} \\ 
    \bottomrule
   \end{tabular}
    }
   \footnotesize{ \\
    \A Correlated lines are weak.
    \B Lines of \ce{^{34}S^{13}CS} and \ce{^{34}S^{12}CS} are partially resolved.}
  \label{tab:FragmentationTable_1}%
  \end{minipage}
\end{table}

Strong correlations and molecule-specific rotational lines were observed between abundant \ce{CS2} isotopologues and their atomic and molecular fragments. Relevant sections in Fig.\ \ref{fgr:CorrelationTable} appear as triangular correlation blocks \circledd{a} for \ce{^{12}C^{32}S2}, \circledd{b} for \ce{^{33}S^{12}C^{32}S}, and \circledd{c} for \ce{^{34}S^{12}C^{32}S}. Correlation between the \ce{^{12}C^{34}S2} isotopologue and its \ce{^{12}C^{34}S} and \ce{^{34}S} fragments, marked \circledd{d}, were weak but could be traced to isotopologue-specific lines. The \ce{CS2} dimer showed clear correlation to \ce{C2S2}, \ce{CS2} and \ce{S2} fragments, marked \circledd{e}.

\begin{figure}[ht]
\centering
\includegraphics[width=6.4cm]{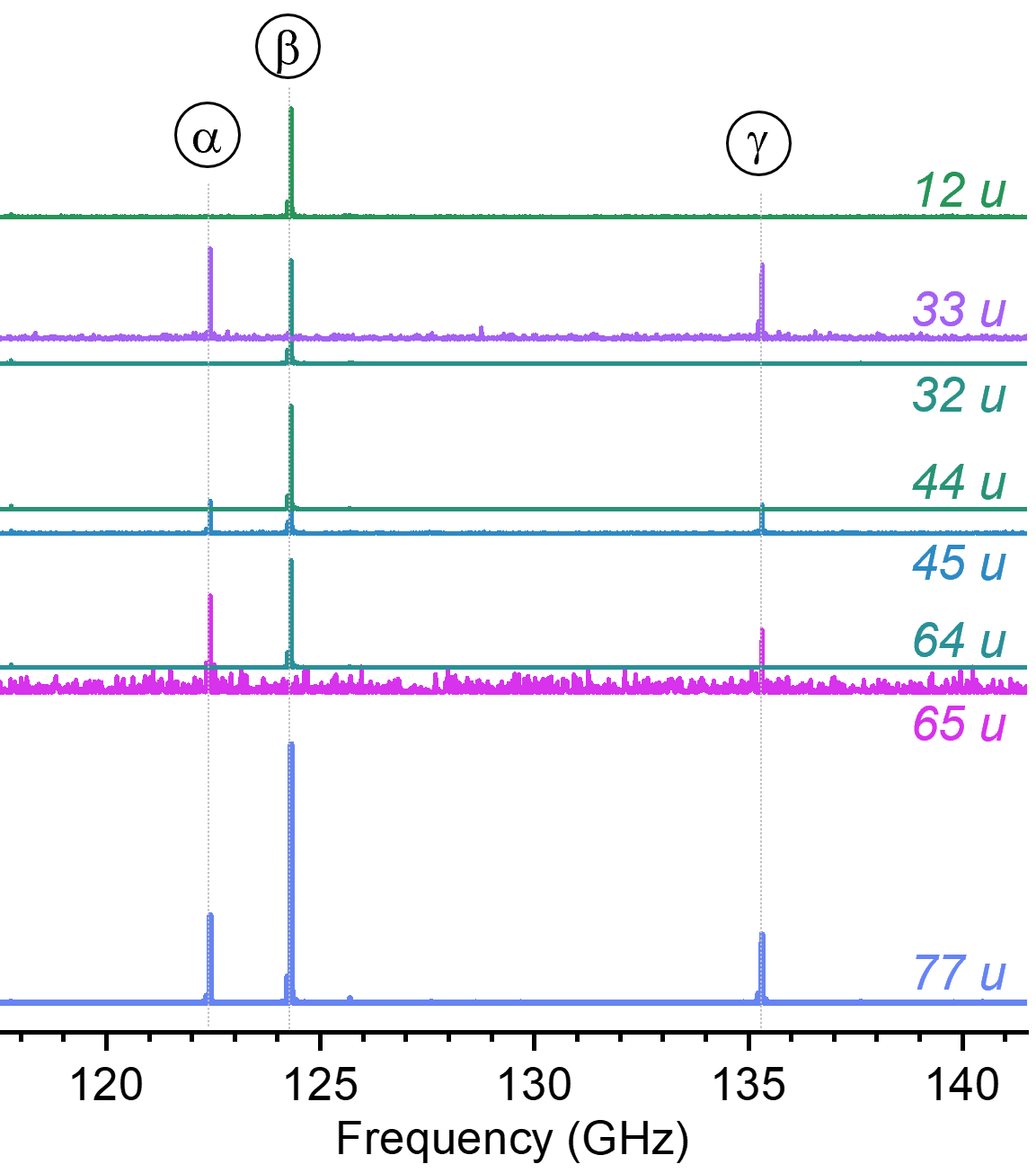}
  \caption[Caption for fgr:LineComparison]{\small Line positions for $J = 8 \leftrightarrow 10$
           (\circledd{$\alpha$},\circledd{$\beta$}) and $J = 9 \leftrightarrow 11$ (\circledd{$\gamma$})
           Raman transitions in mass channel 77 u and possible fragment channels.}
\label{fgr:LineComparison}
\end{figure}

Fig.\ \ref{fgr:LineComparison} illustrates the need to compare molecule-specific line positions for the 77 u mass channel and its fragments. This mass channel contained signals for two isotopologues with similar abundance, \ce{^{13}C^{32}S2} and \ce{^{33}S^{12}C^{32}S}. Lines marked \circledd{$\alpha$} and \circledd{$\gamma$} are unique fingerprints for the \ce{^{33}S^{12}C^{32}S} isotopologue and their presence in the 33 u, 45 u, and 65 u fragment channels allowed the unambiguous assignment of respective fragmentation channels into \ce{^{33}S}, \ce{^{33}S^{12}C} and \ce{^{33}S^{32}S} cationic fragments (see Fig.\ \ref{fgr:CorrelationTable} \circledd{b}). The line marked \circledd{$\beta$} identifies the \ce{^{13}C^{32}S2} isotopologue but is not a unique fingerprint because the more abundant \ce{^{12}C^{32}S2} isotopologue has undistinguishable inertial moments and rotational transition frequencies. The presence of line \circledd{$\beta$} in mass channels 12 u, 32 u, 44 u, and 64 u can therefore not be used to prove the existence of corresponding fragmentation channels. The highly correlated signal block Fig.\ \ref{fgr:CorrelationTable} \circledd{x} is therefore regarded as spurious. The \ce{^{12}C^{32}S2} isotopologue cannot fragment into mass 45 u and the presence of line \circledd{$\beta$} in this mass channel is evidence for fragmentation of \ce{^{13}C^{32}S2} into \ce{^{13}C^{32}S}. The interpretation of signal blocks Fig.\ \ref{fgr:CorrelationTable} \circledd{y} and \circledd{z} required similar considerations and only allowed the assignment of three parent-fragment correlations for the \ce{^{34}S^{12}C^{32}S} isotopologue.

Most of the 28 assigned parent-fragment correlations represent expected fragmentation channels. But the asymmetric fragmentation of the \ce{CS2} dimer cluster into \ce{C2S2} (88 u) and \ce{S2} (64 u) cations came as a surprise. These fragmentation channels require the breaking and formation of covalent bonds, which are much more stable than the intermolecular cluster bonds. DFT calculations\footnote{Nudged elastic band and transition-state optimization at PBEh-3c level, as implemented in ORCA \cite{Neese2012,Neese2018}.} identified possible minimum energy pathways for these cationic fragmentation channels and predicted a considerable energy barrier of 321 kJ/mol for the lowest energy transition state with respect to the vertical ionization energy. The fragment formation energies were also high, with 264 kJ/mol and 323 kJ/mol for fragmentation into \ce{S2 + C2S2^{+}} and \ce{C2S2 + S2^{+}}, respectively. These fragmentation processes require above-threshold ionization (ATI) with absorption of a third UV photon from the photoionization laser beam.

It is possible that excited state dynamics in the neutral cluster, occurring within the ionization pulse duration, increase the yield of ATI and explain the prominence of asymmetric fragmentation. The dynamics in the $^1\Sigma_u^+$ state of \ce{CS2} monomer were studied by time-resolved mass spectrometry \cite{Baronavski1994,Farmanara1999}, time-resolved photoelectron spectroscopy \cite{Townsend2006,Bisgaard2009,Fuji2011,Spesyvtsev2015,Horio2017,Smith2018,Warne2021,Karashima2021,Gabalski2023} and time-resolved electron diffraction \cite{Gabalski2022,Razmus2022}. All studies revealed strong vibrational activity in stretching and bending modes and sub-picosecond appearance of atomic sulfur fragments. We therefore suggest that dynamics in the neutral excited state lead to strong deformation of the dimer geometry, suppressing (increasing) the ionization cross sections into lower (higher) ionic states. Strong electronic coupling with a state of  ($5\sigma_u)^1(7\sigma_g)^1$ character, as described by Horio et al.\ \cite{Horio2017}, may also enhance ATI.

Not just the presence, but also the absence of correlation carries most meaningful information. A significant ion signal was observed for carbon trisulfide (\ce{CS3}, 108 u) but was not correlated to the dimer. \ce{CS3} was previously reported as a stable molecule in the neutral, cationic and anionic state \cite{Suelzle1990,Ma2009,Maity2013} and neutral \ce{CS3} can be generated via collisions of \ce{CS2} with sulfur radicals \cite{Ma2009,Gao2011}. We expected a corresponding reaction channel in the cation following \ce{CS2 ->[h\nu] CS + S^{+}} photofragmentation of a monomer in the dimer. This formation channel, however, must be rejected due to the absence of correlation between  \ce{CS3} and \ce{CS2} dimer signals. \ce{CS3} must therefore be formed by another mechanism, either at thermal equilibrium in the sample or via a very efficient pathway from larger cluster cations. We favor the former hypothesis because the amount of larger clusters in our molecular beam was very small: observed trimer signals were about 200-fold smaller than those of the dimer and 115-fold smaller than those of the \ce{CS3} signals.

\ce{CS2} is a simple molecule and the investigated molecular beam contains only small clusters. For such simple samples, it is tempting to assign observed signals based on preconceived notions about the behavior of each sample component. The assignment of 28 fragmentation channels in our data should caution us against such an approach and illustrates an emerging complexity that can only be addressed with correlated spectroscopy and computer-aided correlation analysis. The method of correlated rotational spectroscopy is readily extendable to larger and more complex systems. The median resolution of rotational spectra in the data set presented here is 18 MHz, on par with the world's highest resolution FTIR spectra\cite{Albert2018} and easily sufficient to resolve complex molecular spectra. Current experiments are therefore not limited by fundamental physical or technological limitations, but only by our ability to collect large data sets within the budget and infrastructure constraints of a small University research laboratory.




Beyond the analysis of molecular clusters, the CRASY method will prove useful for the analysis of other molecular samples that are inherently heterogeneous. Most samples, whether natural or synthetic, are inherently heterogeneous and only the most abundant sample components are routinely purified and characterized. Modern molecular sciences therefore have a blind spot for samples where purification is difficult (e.g., molecular isomers or isotopologues) or where purification is impossible (e.g., neutral clusters or reactive species). Correlated spectroscopy, and the CRASY method in particular, will play an important role to observe previously unseen molecular species and characterize their properties and transformations.

\baselineskip12pt       
\bibliography{CS2_dimer_ARXIV_Version}
\bibliographystyle{h-physrev}

\section*{Acknowledgments}
We acknowledge funding support from the Samsung Science and Technology Foundation, Grant No.\ SSTF-BA2001-08.

\end{document}